\documentclass[12pt]{article}

\usepackage{amssymb}
\usepackage{epsfig}
\usepackage{amsmath}

\begin{document}

\title{Correlations among elastic and inelastic cross-sections and slope parameter}

\author{A. P. Samokhin \\
\textit{A.A. Logunov Institute for High Energy Physics}\\
\textit{of NRC ``Kurchatov Institute''}\\
\textit{Protvino, 142281, Russian Federation}}

\date{}

\maketitle

\begin{abstract}
We discuss the unitarity motivated relations among the elastic cross-section, slope parameter and inelastic cross-section of the 
high energy \textit{pp} interaction. In particular, the MacDowell-Martin unitarity bound is written down in another form to make a relation 
between the elastic and inelastic quantities more transparent. On the basis of an unitarity motivated relation we argue that the growth with energy 
of the elastic to total cross-section ratio is a consequence of the increasing with energy of the \textit{inelastic interaction intensity}.
The latter circumstance is an underlying reason for the acceleration of the slope parameter growth, for the slowing of the growth of the elastic 
to total cross-section ratio and for other interesting phenomena, which are observed in the TeV energy range. All of this confirms the old idea that 
the elastic scattering is a shadow of the particle production processes.
\end{abstract}


\section{Introduction}
A growth with energy of the \textit{pp} total cross-section $ \sigma_{\mathrm{tot}}(s)=\sigma_{\mathrm{el}}(s)+\sigma_{\mathrm{inel}}(s)$ is due
to that of the elastic $\sigma_{\mathrm{el}}(s)$ and the inelastic $\sigma_{\mathrm{inel}}(s)$ cross-sections [1--7]. If the growth
of $\sigma_{\mathrm{inel}}(s)= \sum\limits_{n,\,\mathrm{inel}}^{N(s)} \sigma_{n}(s)$ can be formally attributed to a huge number of open inelastic 
channels $N(s)$ [8], the underlying reasons of the $\sigma_{\mathrm{el}}(s)$ growth are unknown. On the other hand, the unitarity condition relates 
the properties of the elastic scattering amplitude with the contribution from the inelastic channels and the elastic scattering can therefore be 
considered as a shadow of the particle production processes [9]. In other words, due to unitarity there are some correlations between behaviour
of the characteristics of the elastic and inelastic scattering. Indeed, the MacDowell-Martin unitarity bound [10] gives such a relation among 
the total cross-section, the elastic cross-section and the slope parameter of the imaginary part of the elastic scattering amplitude
\begin{equation}
B_{\mathrm{I}}(s) \equiv 2[\frac{d}{dt}\ln|\mathrm{Im}\,T(s,t)|]_{t=0} \geq \frac{\sigma_{\mathrm{tot}}^{2}(s)}{18\pi\sigma_{\mathrm{el}}(s)} .
\end{equation}
In the present note we rewrite this inequality in another form to make a relation between the elastic and inelastic quantities more transparent.

According to the optical theorem (which is a consequence of the unitarity condition) the elastic differential cross-section at zero value of
the square of the four-momentum transfer, $t$, is related to the total cross-section as
\begin{equation}
\frac{d\sigma}{dt}\arrowvert_{t=0}=\frac{\sigma_{\mathrm{tot}}^{2}(s)(1+\rho^{2}(s))}{16\pi},\,\,\,
\rho(s)=\frac{\mathrm{Re}\,T(s,0)}{\mathrm{Im}\,T(s,0)}.
\end{equation}
The slope of the forward diffraction peak, $ B(s)$, and $ (d\sigma/dt)_{t=0}$ are determined experimentally by extrapolation 
of the nuclear elastic scattering differential cross-section data at small values of $t$ to the forward direction $ t=0$ using the exponential form
\begin{equation}
\frac{d\sigma}{dt}=\frac{d\sigma}{dt}\arrowvert_{t=0}\exp(Bt).
\end{equation}
The experimental value of $ \sigma_{\mathrm{tot}}(s)$ is then calculated from Eq. (2) (the ratio of the real to the imaginary part of the 
elastic scattering amplitude in the forward direction, $ \rho(s)$, is taken in this method from the dispersion relations or from global model 
extrapolations). If the local 
slope parameter, $ B(s,t)=d\ln(d\sigma/dt)/dt$, is approximately equal to $ B(s)$ in the essential for the value of integral 
$\sigma_{\mathrm{el}}(s)=\int dt(d\sigma/dt) $ region $0\leq |t|\leq |t_{0}| $, where $ |t_{0}|\approx 0.4\,\mathrm{GeV}^{2}$, the 
elastic cross-section is given by the following formula [11]
\begin{equation}
\sigma_{\mathrm{el}}(s)=\frac{\sigma_{\mathrm{tot}}^{2}(s)(1+\rho^{2}(s))}{16\pi B(s)}.
\end{equation}
At the ISR energies this formula gives a slightly underestimated value for $ \sigma_{\mathrm{el}}(s)$ because the local slope decreases noticeably
with $|t|$ in the $0\leq |t|\leq 0.4\,\mathrm{GeV}^{2} $ range [12], but beyond the ISR energies the relation (4) is practically exact [2--7]. 
The luminosity-independent measurements at 7 TeV [4] give much the same values for $\sigma_{\mathrm{el}}(s)$, $\sigma_{\mathrm{inel}}(s)$ and 
$\sigma_{\mathrm{tot}}(s)$ as the discussed above method and confirm therefore the validity of formula (4). At the LHC energies the local slope
reveals a trend to increase with $|t|$ at $ |t|\gtrsim 0.2\, \mathrm{GeV}^{2}$ [13] due to the nearness of the dip structure of the differential 
cross-section. For that reason, beyond the LHC energies the formula (4) will give a somewhat overestimated value for $ \sigma_{\mathrm{el}}(s)$.
So, in the $10^{2} \lesssim \sqrt{s} \lesssim 5*10^{4}\,\mathrm{GeV}$ energy range Eq. (4) can be considered as a practically exact relation. 
We use it to see the connections between the elastic and inelastic quantities.
Let us note that the MacDowell-Martin bound is close to
Eq. (4) because according to the experimental data $ \rho^{2}(s)\ll1$ and $ B_{\mathrm{I}}(s)\approx B(s)$.

It is an astonishing experimental fact that the elastic cross-section grows faster than the total cross-section, that is the ratio 
$\sigma_{\mathrm{el}}(s)/\sigma_{\mathrm{tot}}(s) $ increases with energy [14, 2--7]. According to Eq. (4) this growth is the same as the
$\sigma_{\mathrm{tot}}(s)/B(s)$ growth (the impact of the $(1+\rho^{2}(s)) $ factor is negligible), which can be interpreted as an increase 
of the interaction intensity in the central part of the interaction region in the impact parameter space [15--17]. Indeed, if the cross-sections
$\sigma_{\mathrm{el}}(s)$, $\sigma_{\mathrm{inel}}(s)$ grow only due to the increasing with energy of the radius of the interaction region $R(s)$, i.e.
$\sigma_{\mathrm{el}},\, \sigma_{\mathrm{inel}} \sim R^{2}$, then the ratios $\sigma_{\mathrm{el}}/B,\, \sigma_{\mathrm{inel}}/B$ are 
energy-independent [1], since $B(s)=0.5 R^{2}(s)$. Hence, the energy growth of the $\sigma_{\mathrm{el}}/B,\, \sigma_{\mathrm{inel}}/B$ ratios has
a dynamical, non geometrical nature and will be referred to as the increasing of the elastic and inelastic interaction intensity respectively. 

From $\sigma_{\mathrm{tot}}=\sigma_{\mathrm{el}}+\sigma_{\mathrm{inel}}$ it is evident that the $\sigma_{\mathrm{el}}/\sigma_{\mathrm{tot}}$ growth
is equivalent to the decreasing of the $\sigma_{\mathrm{inel}}/\sigma_{\mathrm{tot}}$ and $\sigma_{\mathrm{inel}}/\sigma_{\mathrm{el}}$ 
ratios. However, as can be seen from the experimental data [1--7], the ratio $\sigma_{\mathrm{inel}}(s)/B(s)$ is an 
increasing function of energy. It means that the inelastic cross-section grows not only due to the growth of the radius of the interaction 
region but also due to the increasing of the inelastic interaction intensity in the expanding with energy
central part of the interaction region [18--24]. As we will see, the relation (4) enables to get the ratios 
$\sigma_{\mathrm{el}}/B$, $\sigma_{\mathrm{el}}/\sigma_{\mathrm{tot}}$ and $\sigma_{\mathrm{el}}/\sigma_{\mathrm{inel}}$ in the form of increasing 
functions of ratio $(\sigma_{\mathrm{inel}}/B)$. Therefore, the growth of these ratios with energy is a consequence (a ``shadow'') of the 
increasing of the inelastic interaction intensity.

According to the experimental data [1--7] the difference $(\sigma_{\mathrm{inel}}-\sigma_{\mathrm{el}})=(\sigma_{\mathrm{tot}}-2\sigma_{\mathrm{el}})$
monotonically increases with energy (see Fig. 1). Due to the $\sigma_{\mathrm{el}}/\sigma_{\mathrm{tot}}$ growth the ratios of this difference to the 
$\sigma_{\mathrm{el}}$, $\sigma_{\mathrm{inel}}$ and $\sigma_{\mathrm{tot}}$ decrease with energy but the ratio 
$(\sigma_{\mathrm{inel}}-\sigma_{\mathrm{el}})/B$, as can be seen from Eq. (4), grows with energy up to $\sim$ 3 TeV (where 
$\sigma_{\mathrm{el}}/\sigma_{\mathrm{tot}}=0.25$ [7]), reaches its maximum value and then decreases with energy. 
There is evidence that approximately at this energy the slope $B(s)$ begins to accelerate its growth [6, 7] (the discussions of this new phenomenon
in the context of different models can be found in Refs. [25--31]). As can be seen from Ref. [7], the curvature of the 
$\sigma_{\mathrm{el}}/\sigma_{\mathrm{tot}}$ growth changes its sign from positive to negative also in this energy range [16], and therefore
the ratio $\sigma_{\mathrm{el}}/\sigma_{\mathrm{tot}}$ slows down its growth. As we will see, the unitarity motivated Eq. (4) enables to consider
all these phenomena, as well as the $\sigma_{\mathrm{el}}/\sigma_{\mathrm{tot}}$ growth itself, as a consequence of the increasing of the intensity
of the inelastic interaction $(\sigma_{\mathrm{inel}}/B)$.

In Section 2 we give a few new forms of the MacDowell-Martin unitarity bound, which relate the elastic and inelastic quantities. 
The consequences of Eq. (4) are studied in Section 3, where, in addition, we discuss some phenomenological arguments in favour of
the shadow origin of the $\sigma_{\mathrm{el}}(s)$ growth. A brief summary and discussion are given in Section 4.
\section{The MacDowell-Martin bound}
The MacDowell-Martin unitarity bound (1) can be written as
\begin{equation}
\sigma_{\mathrm{tot}}^{2}(s)\leq\tilde{\beta}(s)\sigma_{\mathrm{el}}(s),\,\,\,\tilde{\beta}(s)\equiv 18 \pi B_{\mathrm{I}}(s).
\end{equation}
Taking into account a relation $\sigma_{\mathrm{el}}=(\sigma_{\mathrm{tot}}-\sigma_{\mathrm{inel}})$, we can rewrite Eq. (5) in the following
equivalent form
\begin{equation}
\sigma_{\mathrm{inel}} \leq \sigma_{\mathrm{tot}} (1- \frac{\sigma_{\mathrm{tot}}}{\tilde{\beta}}) \leq \frac{\tilde{\beta}}{4}.
\end{equation}
The last inequality in Eq. (6) is due to the obvious inequality $x(1-x)\leq1/4$. Exactly the same arguments were used in Ref. [32] to obtain
an absolute upper bound
\begin{equation}
\sigma_{\mathrm{inel}} \leq \sigma_{\mathrm{tot}} (1- \frac{\sigma_{\mathrm{tot}}}{\sigma_{\mathrm{max}}}) \leq \frac{\sigma_{\mathrm{max}}}{4},\,\,\,
\sigma_{\mathrm{max}}\equiv\frac{4\pi}{t_{0}}\,\ln^{2}(\frac{s}{s_{0}}).
\end{equation}
\begin{figure}[t]
\centering
\includegraphics[height=8.5cm]{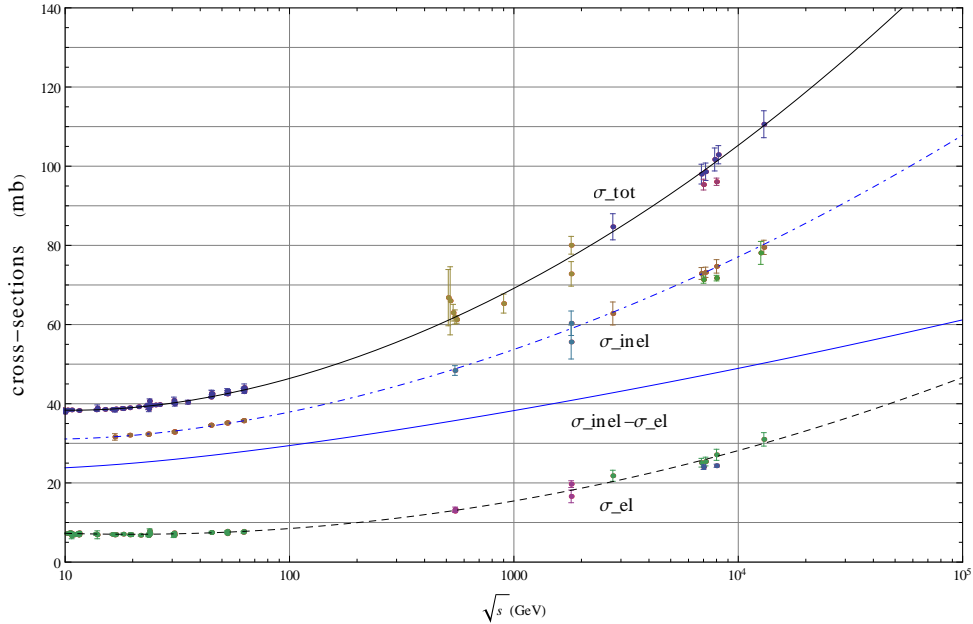}
\caption{The elastic ($\sigma_{\mathrm{el}}$), inelastic ($\sigma_{\mathrm{inel}}$), total ($\sigma_{\mathrm{tot}}$) cross-section for \textit{pp}
collisions as a function of $\sqrt{s}$ [1--7], including the \textit{$\bar{p}$p} data at $\sqrt{s}=546, 900, 1800$ GeV [14]. The continuous
black line is a fit of the total cross-section data by the COMPETE collaboration [33]. The dashed line is a fit of the elastic cross-section 
data by the TOTEM collaboration [7]. The dash-dotted and blue lines refer to the $\sigma_{\mathrm{inel}}=(\sigma_{\mathrm{tot}}-\sigma_{\mathrm{el}})$ 
and $(\sigma_{\mathrm{inel}}-\sigma_{\mathrm{el}})=(\sigma_{\mathrm{tot}}-2\,\sigma_{\mathrm{el}})$ respectively and are obtained as the differences 
between the $\sigma_{\mathrm{tot}}$ and $\sigma_{\mathrm{el}}$ fits.}  
\end{figure}
\begin{figure}[t]
\centering
\includegraphics[height=8.5cm]{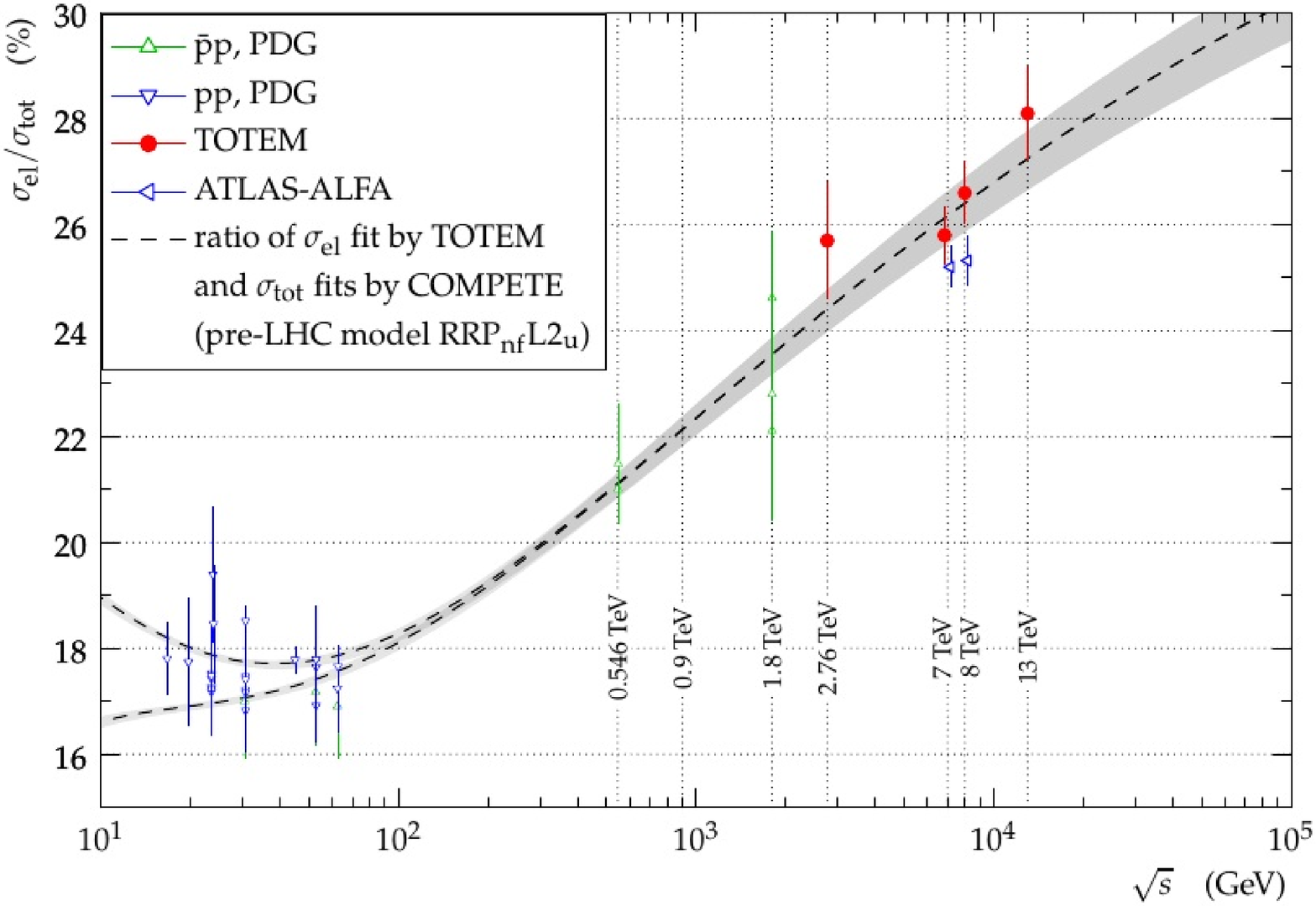}
\caption{The elastic to total cross-section ratio for \textit{pp} and \textit{$\bar{p}$p} collisions as a function of energy $\sqrt{s}$ [7].} 
\end{figure}
\begin{figure}[t]
\centering
\includegraphics[height=8.5cm]{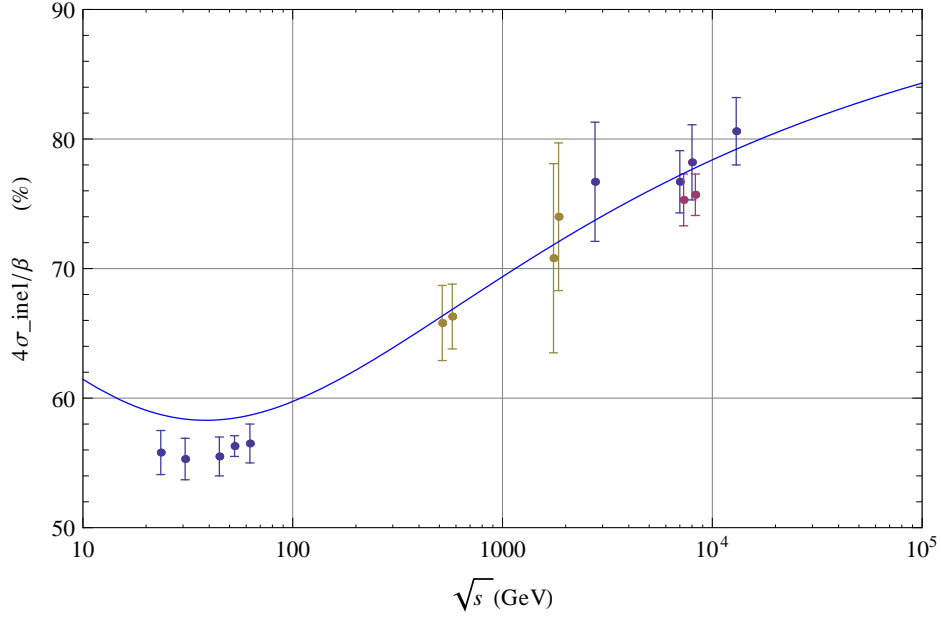}
\caption{The intensity of the inelastic \textit{pp} interaction, $(4\sigma_{\mathrm{inel}}/\beta)$, as a function of energy $\sqrt{s}$ (the experimental 
data are from Refs. [1--7]),
including the \textit{$\bar{p}$p} data at $\sqrt{s}=546$ and 1800 GeV [14]. The line represents the function 
$(4\sigma_{\mathrm{el}}\sigma_{\mathrm{inel}}/\sigma_{\mathrm{tot}}^{2})$, where the fits by the COMPETE collaboration [33] and TOTEM collaboration
[7] are used for $\sigma_{\mathrm{tot}}$ and $\sigma_{\mathrm{el}}$ respectively.}
\end{figure}

The inequality (5) can be written down also as
\begin{equation}
\frac{\sigma_{\mathrm{tot}}^{2}}{\tilde{\beta}^{2}}\,\, \leq \,\,\frac{\sigma_{\mathrm{el}}}{\tilde{\beta}}\,\, \leq\,\, 
\frac{\sigma_{\mathrm{el}}^{2}}{\sigma_{\mathrm{tot}}^{2}}.
\end{equation}
From the left-hand side (lhs) of Eq. (8) we have the following form of the MacDowell-Martin bound
\begin{equation}
\frac{(\sigma_{\mathrm{inel}}-\sigma_{\mathrm{el}})}{\tilde{\beta}}=\frac{(\sigma_{\mathrm{tot}}-2\,\sigma_{\mathrm{el}})}{\tilde{\beta}}\,\,\leq\,\,
\frac{\sigma_{\mathrm{tot}}}{\tilde{\beta}}\,(1-2\,\frac{\sigma_{\mathrm{tot}}}{\tilde{\beta}})\,\,\leq\,\,\frac{1}{8}.
\end{equation}
If $(\sigma_{\mathrm{inel}}-\sigma_{\mathrm{el}})\geq0$ then from $(1/\tilde{\beta})\leq(\sigma_{\mathrm{el}}/\sigma_{\mathrm{tot}}^{2})$ it follows
also
\begin{equation}
\frac{(\sigma_{\mathrm{inel}}-\sigma_{\mathrm{el}})}{\tilde{\beta}}\,\,\leq\,\,
\frac{\sigma_{\mathrm{el}}}{\sigma_{\mathrm{tot}}}\,(1-2\,\frac{\sigma_{\mathrm{el}}}{\sigma_{\mathrm{tot}}})\,\,\leq\,\,\frac{1}{8}.
\end{equation}
The lhs of Eq. (8) gives the following interesting inequalities
\begin{equation}
\frac{\sigma_{\mathrm{inel}}}{\tilde{\beta}}\,\leq\,\sqrt{\frac{\sigma_{\mathrm{el}}}{\tilde{\beta}}}(1-\sqrt{\frac{\sigma_{\mathrm{el}}}
{\tilde{\beta}}}\,),
\,\,\,\,\frac{1}{2}(1-\sqrt{1-4\,\frac{\sigma_{\mathrm{inel}}}{\tilde{\beta}}}\,)\,\leq\,\frac{\sigma_{\mathrm{tot}}}{\tilde{\beta}}\,\leq\,
\frac{\sigma_{\mathrm{el}}}{\sigma_{\mathrm{tot}}},
\end{equation}
\begin{equation}
\frac{1}{2}[(1-2\,\frac{\sigma_{\mathrm{inel}}}{\tilde{\beta}}\,)-\sqrt{1-4\,\frac{\sigma_{\mathrm{inel}}}{\tilde{\beta}}}\,]\,\leq\,
\frac{\sigma_{\mathrm{el}}}{\tilde{\beta}}\,\leq\,
\frac{\sigma_{\mathrm{el}}^{2}}{\sigma_{\mathrm{tot}}^{2}},
\end{equation}
which relate the ratios $\sigma_{\mathrm{el}}/\tilde{\beta}$ and $\sigma_{\mathrm{inel}}/\tilde{\beta}$.
Let us note that all these inequalities are the consequences of the unitarity condition and have the same status as original MacDowell-Martin bound (1).
As well as Eq. (1), these inequalities are near to the saturation.
\section{Correlations among $\sigma_{\mathrm{el}}$, $\sigma_{\mathrm{inel}}$, 
$\sigma_{\mathrm{tot}}$ and $ B$}
Let us rewrite Eq. (4) in the following form
\begin{equation}
\frac{\sigma_{\mathrm{el}}}{\beta}=\frac{\sigma_{\mathrm{tot}}^{2}}{\beta^{2}}=\frac{\sigma_{\mathrm{el}}^{2}}{\sigma_{\mathrm{tot}}^{2}}\,,\,\,\,
\beta \equiv \frac{16\pi B}{(1+\rho^{2})} \approx 16\pi B.
\end{equation}
Then, it is evident that
\begin{equation}
\frac{\sigma_{\mathrm{inel}}}{\beta}=\frac{\sigma_{\mathrm{el}}\,\sigma_{\mathrm{inel}}}{\sigma_{\mathrm{tot}}^{2}}=\frac{\sigma_{\mathrm{tot}}}{\beta}(1-\frac{\sigma_{\mathrm{tot}}}{\beta})=
\frac{\sigma_{\mathrm{el}}}{\sigma_{\mathrm{tot}}}(1-\frac{\sigma_{\mathrm{el}}}{\sigma_{\mathrm{tot}}})\leq \frac{1}{4}.
\end{equation}
So, since $\sigma_{\mathrm{el}}/\sigma_{\mathrm{tot}}<1/2$, the growth of $\sigma_{\mathrm{el}}/\sigma_{\mathrm{tot}}$ gives the growth of
$\sigma_{\mathrm{inel}}/{\beta}$ and vice versa (see Fig. 2 and Fig. 3)
\begin{equation}
\frac{\sigma_{\mathrm{el}}}{\sigma_{\mathrm{tot}}}=\frac{\sigma_{\mathrm{tot}}}{\beta}=\frac{1}{2}(1-\sqrt{1-4x}),\,\,\,
x\equiv\frac{\sigma_{\mathrm{inel}}}{\beta}.
\end{equation}
Therefore, the growth of $\sigma_{\mathrm{el}}/\sigma_{\mathrm{tot}}$ with energy can be considered as a consequence of the 
$\sigma_{\mathrm{inel}}/\beta$ growth, i.e. as an enforced behaviour (as a ``shadow'') due to the increasing of the intensity of the inelastic 
interaction. The same is true for the growth with energy of the elastic interaction intensity $\sigma_{\mathrm{el}}/\beta$ and ratio of the 
intensities of the elastic and inelastic interaction $\sigma_{\mathrm{el}}/\sigma_{\mathrm{inel}}$
\begin{equation}
\frac{\sigma_{\mathrm{el}}}{\beta}=\frac{1}{2}(1-2x-\sqrt{1-4x}),\,\,\,\frac{\sigma_{\mathrm{el}}}{\sigma_{\mathrm{inel}}}=
\frac{1-\sqrt{1-4x}}{1+\sqrt{1-4x}},
\end{equation}
because the functions in the rhs of Eq. (16) monotonically increase with $x$.
\begin{figure}[t]
\centering
\includegraphics[height=8.5cm]{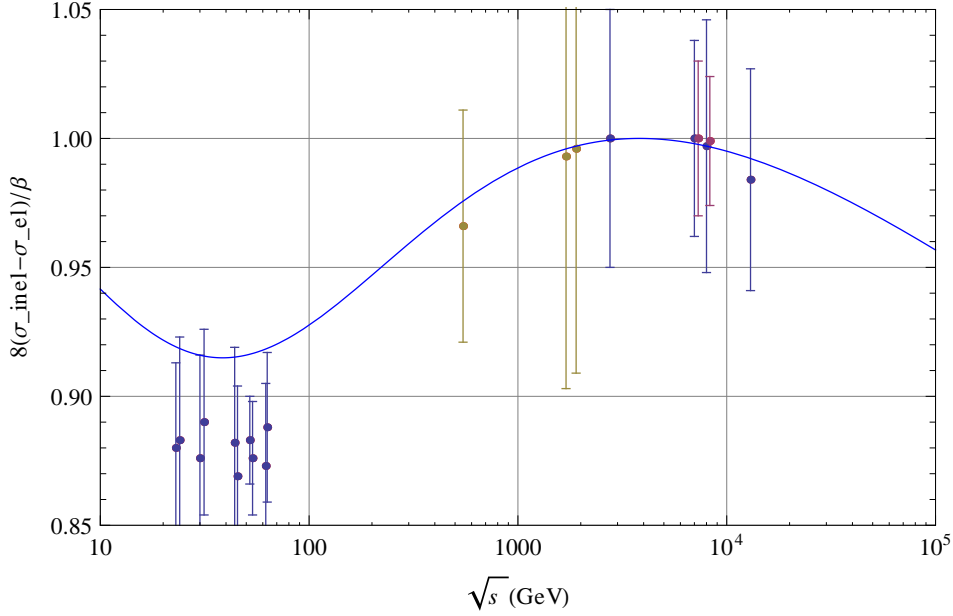}
\caption{The quantity $8(\sigma_{\mathrm{inel}}-\sigma_{\mathrm{el}})/\beta$ for \textit{pp} collisions as a function of energy $\sqrt{s}$ (the experimental 
data are from Refs. [1--7]), including the \textit{$\bar{p}$p} data at $\sqrt{s}=546$ and 1800 GeV [14]. The line represents the function 
$8\sigma_{\mathrm{el}}(\sigma_{\mathrm{tot}}-2\sigma_{\mathrm{el}})/\sigma_{\mathrm{tot}}^{2}$, where the fits by the COMPETE collaboration [33] 
and TOTEM collaboration [7] are used for $\sigma_{\mathrm{tot}}$ and $\sigma_{\mathrm{el}}$ respectively.}
\end{figure}
\begin{figure}[t]
\centering
\includegraphics[height=8.5cm]{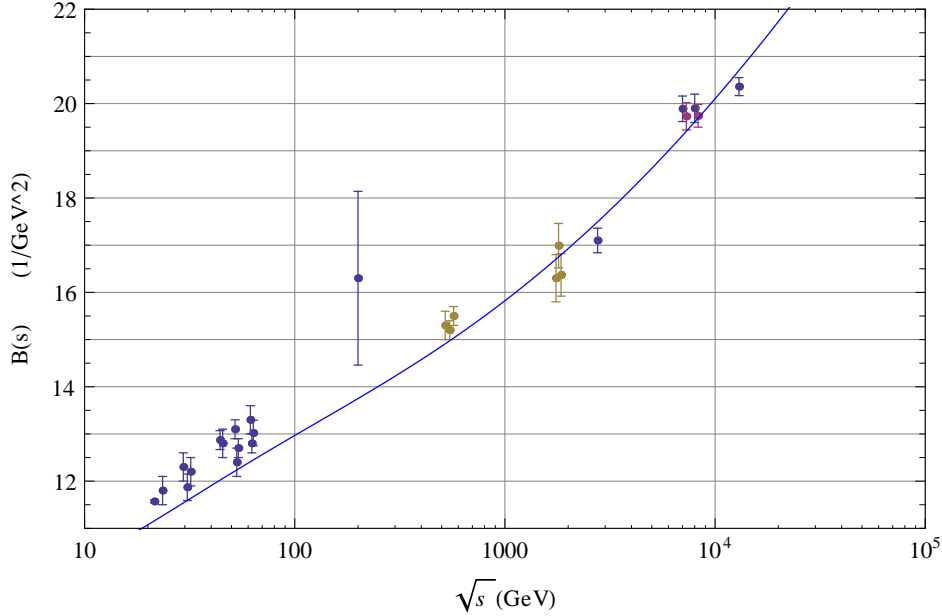}
\caption{The nuclear slope $B(s)$ for \textit{pp} elastic scattering as a function of energy $\sqrt{s}$ (the experimental 
data are from Refs. [1--7]), including the \textit{$\bar{p}$p} data at $\sqrt{s}=546$ and 1800 GeV [14]. The line represents the function 
$(\sigma_{\mathrm{tot}}^{2}/16 \pi \sigma_{\mathrm{el}})$, where the fits by the COMPETE collaboration [33] and TOTEM collaboration
[7] are used for $\sigma_{\mathrm{tot}}$ and $\sigma_{\mathrm{el}}$ respectively.} 
\end{figure}

According to the experimental data [1--7] the difference $(\sigma_{\mathrm{inel}}-\sigma_{\mathrm{el}})=(\sigma_{\mathrm{tot}}-2\sigma_{\mathrm{el}})$
increases with energy (see Fig. 1).
It is evident that the ratios of this difference to the $\sigma_{\mathrm{el}}$, $\sigma_{\mathrm{inel}}$ and $\sigma_{\mathrm{tot}}$ decrease
with energy but from Eq. (13) we see that the ratio
\begin{equation}
\frac{\sigma_{\mathrm{inel}}-\sigma_{\mathrm{el}}}{\beta}=\frac{\sigma_{\mathrm{tot}}-2\sigma_{\mathrm{el}}}{\beta}=
\frac{\sigma_{\mathrm{el}}}{\sigma_{\mathrm{tot}}}(1-2\frac{\sigma_{\mathrm{el}}}{\sigma_{\mathrm{tot}}})\leq \frac{1}{8}
\end{equation}
increases in the energy range where $\sigma_{\mathrm{el}}/\sigma_{\mathrm{tot}}<0.25$ and decreases with energy when 
$\sigma_{\mathrm{el}}/\sigma_{\mathrm{tot}}>0.25$ (see Fig. 4). So, this ratio changes its energy behaviour at $\sqrt{s}\sim 3 $ TeV where
$\sigma_{\mathrm{el}}/\sigma_{\mathrm{tot}}=0.25$ [7]. Approximately in this energy range
the growth rate of the slope $B(s)$ begins to increase (see Fig. 5) and the curvature of the $\sigma_{\mathrm{el}}(s)/\sigma_{\mathrm{tot}}(s)$ growth 
changes its sign from positive to negative [6,7] (see Fig. 2). All these phenomena are related between each other and, as we will see, are due to the 
$\sigma_{\mathrm{el}}/\sigma_{\mathrm{tot}}$ growth, i.e. the increasing of the intensity of the inelastic interaction (see Fig. 3).

As can be seen from Fig. 1, the cross-section $\sigma_{\mathrm{el}}=(\sigma_{\mathrm{tot}}-\sigma_{\mathrm{inel}})$ 
grows faster than $ \ln (\sqrt{s})$ and 
the difference $(\sigma_{\mathrm{inel}}-\sigma_{\mathrm{el}})=(\sigma_{\mathrm{tot}}-2\sigma_{\mathrm{el}})=
(2\sigma_{\mathrm{inel}}-\sigma_{\mathrm{tot}})$ grows not slower than $ \ln (\sqrt{s})$.
Therefore, from the positivity of $\sigma_{\mathrm{el}}=(\sigma_{\mathrm{tot}}-\sigma_{\mathrm{inel}})$, 
$(\sigma_{\mathrm{inel}}-\sigma_{\mathrm{el}})$ and their derivatives we have
\begin{equation}
0<\sigma_{\mathrm{el}}<0.5\,\sigma_{\mathrm{tot}}<\sigma_{\mathrm{inel}}<\sigma_{\mathrm{tot}},\,\,\,\,\,
0<\sigma_{\mathrm{el}}'<0.5\,\sigma_{\mathrm{tot}}'<\sigma_{\mathrm{inel}}'<\sigma_{\mathrm{tot}}',\,\,\,
\end{equation}
\begin{equation}
\mathrm{and}\,\,\,0<\sigma_{\mathrm{el}}''\leq 0.5\,\sigma_{\mathrm{tot}}''\leq \sigma_{\mathrm{inel}}''<\sigma_{\mathrm{tot}}'',
\end{equation}
where the prime denotes the derivative with respect to $ \ln (\sqrt{s})$. The growth of the ratios $\sigma_{\mathrm{el}}/\sigma_{\mathrm{tot}}$, 
$\sigma_{\mathrm{tot}}/\sigma_{\mathrm{inel}}$, $\sigma_{\mathrm{inel}}/B$ and the growth of $B(s)$ 
\begin{equation}
B(s)=\frac{\sigma_{\mathrm{tot}}^{2}(s)}{16\pi \sigma_{\mathrm{el}}(s)}
\end{equation}
(the impact of the $(1+\rho^{2}(s)) $ factor is negligible and we omit it) give the following chain of inequalities
\begin{equation}
0<(2\,\frac{\sigma_{\mathrm{tot}}'}{\sigma_{\mathrm{tot}}}-\frac{\sigma_{\mathrm{el}}'}{\sigma_{\mathrm{el}}})=
\frac{B'}{B}<\frac{\sigma_{\mathrm{inel}}'}{\sigma_{\mathrm{inel}}}<\frac{\sigma_{\mathrm{tot}}'}{\sigma_{\mathrm{tot}}}<
\frac{\sigma_{\mathrm{el}}'}{\sigma_{\mathrm{el}}}<2\,\frac{\sigma_{\mathrm{tot}}'}{\sigma_{\mathrm{tot}}}.
\end{equation}

The second derivative of the ratio $\sigma_{\mathrm{el}}/\sigma_{\mathrm{tot}}$ with respect to $ \ln (\sqrt{s})$ has the form
\begin{equation}
(\frac{\sigma_{\mathrm{el}}}{\sigma_{\mathrm{tot}}})''=\frac{\sigma_{\mathrm{el}}}{\sigma_{\mathrm{tot}}}
[\frac{1}{\sigma_{\mathrm{el}}}(\sigma_{\mathrm{el}}''-\frac{\sigma_{\mathrm{el}}}{\sigma_{\mathrm{tot}}}\,\sigma_{\mathrm{tot}}'')-2\,
\frac{\sigma_{\mathrm{tot}}'}{\sigma_{\mathrm{tot}}}(\frac{\sigma_{\mathrm{el}}'}{\sigma_{\mathrm{el}}}-
\frac{\sigma_{\mathrm{tot}}'}{\sigma_{\mathrm{tot}}})].
\end{equation}
According to Eqs. (19), (21) only the first term in this relation has a positive sign. Since the $\sigma_{\mathrm{el}}''$, $\sigma_{\mathrm{tot}}''$
are practically constant, the growth of the value of $\sigma_{\mathrm{el}}/\sigma_{\mathrm{tot}}$ reduces the impact of this positive term
and leads to zeroing of $(\sigma_{\mathrm{el}}/\sigma_{\mathrm{tot}})''$ and then to $(\sigma_{\mathrm{el}}/\sigma_{\mathrm{tot}})''<0$. So, due to
the growth of the value of $\sigma_{\mathrm{el}}/\sigma_{\mathrm{tot}}$, i.e. due to the increasing of the intensity of the inelastic interaction,
the curvature of the $\sigma_{\mathrm{el}}(s)/\sigma_{\mathrm{tot}}(s)$ function changes its sign in the TeV energy range and the growth of
$\sigma_{\mathrm{el}}/\sigma_{\mathrm{tot}}$ slows down (see Fig. 2). Let us note that the curvature of the $\sigma_{\mathrm{inel}}(s)/\beta(s)$ 
function changes its sign from positive to negative at some lower energy than that for the $\sigma_{\mathrm{el}}(s)/\sigma_{\mathrm{tot}}(s)$ 
function (see Fig. 2 and Fig. 3), since
\begin{equation}
x''=(\frac{\sigma_{\mathrm{inel}}}{\beta})''=(1-2\,\frac{\sigma_{\mathrm{el}}}{\sigma_{\mathrm{tot}}})(\frac{\sigma_{\mathrm{el}}}
{\sigma_{\mathrm{tot}}})''-2\,((\frac{\sigma_{\mathrm{el}}}{\sigma_{\mathrm{tot}}})')^{2}.
\end{equation}

From Eq. (20) we have
\begin{equation}
\frac{B''}{B}=\frac{1}{\sigma_{\mathrm{el}}}(2\,\frac{\sigma_{\mathrm{el}}}{\sigma_{\mathrm{tot}}}\,\sigma_{\mathrm{tot}}''-\sigma_{\mathrm{el}}'')+
2\,(\frac{\sigma_{\mathrm{el}}'}{\sigma_{\mathrm{el}}}-
\frac{\sigma_{\mathrm{tot}}'}{\sigma_{\mathrm{tot}}})^{2}.
\end{equation}
When the ratio $\sigma_{\mathrm{el}}/\sigma_{\mathrm{tot}}$ is small the second negative term compensates the positive terms in Eq. (24) 
and $B''\approx 0$, i.e. the slope $B(s)$ is approximately a linear function of $\ln (\sqrt{s})$ in this energy range. As for
$(\sigma_{\mathrm{el}}/\sigma_{\mathrm{tot}})''$, the growth of the $\sigma_{\mathrm{el}}/\sigma_{\mathrm{tot}}$ value reduces the impact of
this negative term and in the TeV energy range the growth rate of $B(s)$ increases (see Fig. 5). Due to the factor 2 in the first term of Eq. (24), 
the acceleration of the $B(s)$ growth is begun at lower energy than the value of energy at which the curvature of 
$\sigma_{\mathrm{el}}/\sigma_{\mathrm{tot}}$ changes its sign (see Fig. 2 and Fig. 5). So, we see that the acceleration of the $B(s)$ growth 
in the TeV energy range is a consequence of the increasing of the intensity of the inelastic interaction.
\begin{figure}[t]
\centering
\includegraphics[height=8.5cm]{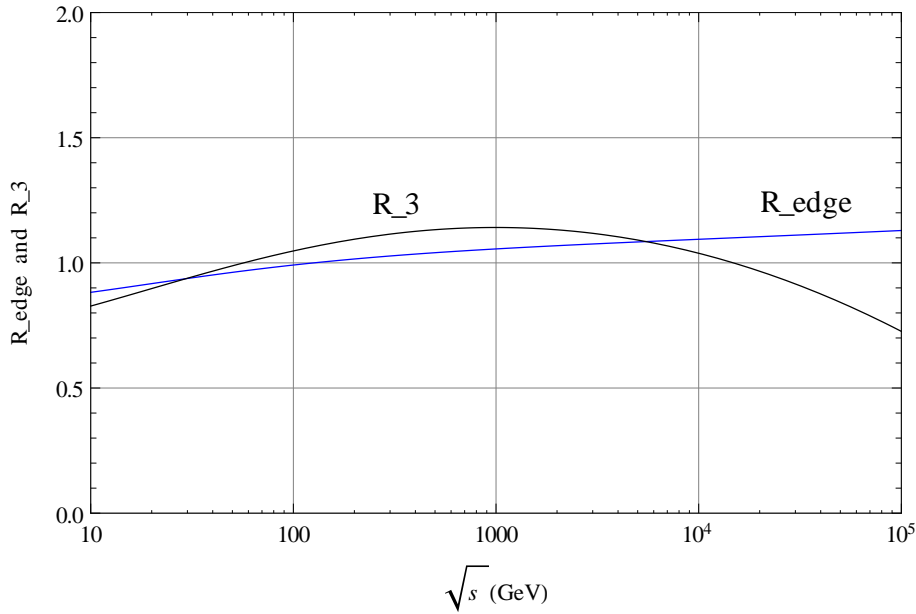}
\caption{The ratio $\mathrm{R}_{\mathrm{edge}}=(\sigma_{\mathrm{inel}}-\sigma_{\mathrm{el}})/\sqrt{(19\,\mathrm{mb})\,\sigma_{\mathrm{tot}}}\,,$  
$\mathrm{R}_{3}=(\sigma_{\mathrm{tot}}-3\,\sigma_{\mathrm{el}})/20\,\mathrm{mb}$ for \textit{pp} collisions as a function of energy $\sqrt{s}$.
Here the fits by the COMPETE collaboration [33] and TOTEM collaboration
[7] are used for $\sigma_{\mathrm{tot}}$ and $\sigma_{\mathrm{el}}$ respectively.} 
\end{figure}

According to the phenomenological arguments of Ref. [34] (see also [26, 35, 36]) the difference $(\sigma_{\mathrm{inel}}-\sigma_{\mathrm{el}})$
increases with energy as $\sqrt{\sigma_{\mathrm{tot}}}$ , i.e.
\begin{equation}
\frac{\sigma_{\mathrm{inel}}-\sigma_{\mathrm{el}}}{\sqrt{\pi \sigma_{\mathrm{tot}}/2}}=
\frac{\sigma_{\mathrm{tot}}-2\sigma_{\mathrm{el}}}{\sqrt{\pi \sigma_{\mathrm{tot}}/2}}=t_{\mathrm{edge}}\,,
\end{equation}
where $t_{\mathrm{edge}}\approx1.1$ fm is an approximately constant width of the ``edge''. We can rewrite this relation in the 
following equivalent forms
\begin{equation}
\sigma_{\mathrm{el}}(s)=\sigma_{\mathrm{inel}}(s)+c-\sqrt{c\,(4\,\sigma_{\mathrm{inel}}(s)+c)}\,,
\end{equation}
\begin{equation}
\sigma_{\mathrm{inel}}(s)=\sigma_{\mathrm{el}}(s)+c+\sqrt{c\,(4\,\sigma_{\mathrm{el}}(s)+c)}\,,
\end{equation}
where $c\approx9.5$ mb is a constant. Since $\sigma_{\mathrm{inel}}(s)>2\,c$, the derivative of Eq. (26) with respect to $ \ln (\sqrt{s})$
\begin{equation}
\sigma_{\mathrm{el}}'(s)=(1-\frac{2c}{\sqrt{c\,(4\,\sigma_{\mathrm{inel}}(s)+c)}})\,\sigma_{\mathrm{inel}}'(s)>\frac{\sigma_{\mathrm{inel}}'(s)}{3}
\end{equation}
is positive and therefore the growth of $ \sigma_{\mathrm{el}}(s)$ is a consequence of the $\sigma_{\mathrm{inel}}(s)$ growth. So, Eq. (26) gives
a direct evidence for the shadow origin of the $\sigma_{\mathrm{el}}(s)$ growth. Since $(\sigma_{\mathrm{inel}}-\sigma_{\mathrm{el}})=
\sigma_{\mathrm{tot}}\,(1-2\,\sigma_{\mathrm{el}}/\sigma_{\mathrm{tot}})$, Eq. (25) can also be written as 
\begin{equation}
\frac{\sigma_{\mathrm{el}}}{\sigma_{\mathrm{tot}}}\approx \frac{1}{2}(1-\sqrt{\frac{2c}{\sigma_{\mathrm{tot}}}}\,).
\end{equation}

In conclusion, let us mention the following purely empirical approximate relation
\begin{equation}
\sigma_{\mathrm{el}}(s)\approx 0.5\,\sigma_{\mathrm{inel}}(s)-10\,\mathrm{mb}\,\,\,\,\, \mathrm{or} \,\,\,\,\, 
\sigma_{\mathrm{tot}}(s)\approx 3\, \sigma_{\mathrm{el}}(s)+20\,\mathrm{mb},
\end{equation}
which is valid in the ISR -- LHC energy range. The relative accuracy of Eqs. (25) -- (30) in the ISR -- LHC energy range is approximately 10 -- 15$\%$
(see Fig. 6).

It is interesting to note that the intensity of the inelastic interaction $(\sigma_{\mathrm{tot}}-\sigma_{\mathrm{el}})/\beta$ monotonically grows
with energy (see Fig. 3), the quantity $(\sigma_{\mathrm{tot}}-2\sigma_{\mathrm{el}})/\beta$ reveals the discussed above properties (see Fig. 4) but
the function
\begin{equation}
\frac{\sigma_{\mathrm{inel}}-2\sigma_{\mathrm{el}}}{\beta}=\frac{\sigma_{\mathrm{tot}}-3\sigma_{\mathrm{el}}}{\beta}=
\frac{\sigma_{\mathrm{el}}}{\sigma_{\mathrm{tot}}}(1-3\frac{\sigma_{\mathrm{el}}}{\sigma_{\mathrm{tot}}})\leq \frac{1}{12}
\end{equation}
monotonically decreases with energy (when the ratio $\sigma_{\mathrm{el}}/\sigma_{\mathrm{tot}}$ grows), since 
$(\sigma_{\mathrm{el}}/\sigma_{\mathrm{tot}})>1/6$ at all energies (see Fig. 2).
\section{Discussion}

The size of the hadron-hadron interaction region at all available energies is comparable with the own size of hadrons [27, 28]. Due to this cause 
we have not been able to construct a self-consistent theory of the hadron-hadron elastic and diffraction scattering from the first QCD principles.
Moreover, even the general qualitative features of the elastic scattering of the extended, composite objects are unknown and due to that, the elastic 
hadron-hadron scattering properties seem puzzling. In contrast, the nondiffraction inelastic processes are an ordinary subject of the QCD studying
and the growth of the intensity of such interactions with energy seems quite natural.

On the other hand, it is well known that due to the unitarity condition the properties of the elastic scattering amplitude are related with that 
of a sum of the inelastic channels and the elastic scattering can therefore be considered as a shadow of the particle production processes.
In present note on the basis of the unitarity motivated relations we have demonstrated the effectiveness of such an approach. 
At first sight such a phenomenon as the growth with energy of the value of the ratio $\sigma_{\mathrm{el}}/\sigma_{\mathrm{inel}}$ (and hence of the
value of the ratio $\sigma_{\mathrm{el}}/\sigma_{\mathrm{tot}}$) seems puzzling but actually turns out to be a simple consequence of the increasing 
of the inelastic interaction intensity $(\sigma_{\mathrm{inel}}/\beta)$. In other words, the growth of the intensity of the elastic interaction, 
$(\sigma_{\mathrm{el}}/\beta)$,
and the growth of the ratio of this intensity to the inelastic interaction intensity, $\sigma_{\mathrm{el}}/\sigma_{\mathrm{inel}}$, are an unitarity
shadow of the increasing of the intensity of the inelastic interaction. The increasing of this fundamental quantity with energy determines the 
changes in the behaviour of the slope and ratio $\sigma_{\mathrm{el}}/\sigma_{\mathrm{tot}}$ in the TeV energy range, where the most sensitive 
quantity, the ratio $ (\sigma_{\mathrm{inel}}-\sigma_{\mathrm{el}})/\beta$, as we have seen, changes its growth to the decreasing.

Another consequence of the increasing of the intensity of the inelastic interaction was discussed in Ref. [23], where it was argued that
growth with energy of the central profile of the probability of an inelastic interaction $P_{\mathrm{inel}}(s,b)$ leads to the formation
of the peripheral profile of the real part $r(s,b)$ of the elastic scattering amplitude in the impact parameter space, if the imaginary part
of the elastic scattering amplitude $a(s,b=0)>0.5$. Such behaviour of the real part of the scattering amplitude is very natural for 
the elastic scattering.

The correlated growth of the slope and total cross-section leads also to a stationary point in the energy behaviour of the differential 
cross-section in the forward diffraction cone, which is observed in the high energy \textit{pp} scattering [37]. Let us note, by the way, that 
the use of a mean value of the local slope in the range $0\leq |t|\leq 0.21\,\mathrm{GeV}^{2}$ instead $B(s)$ [37] significantly increases 
the accuracy of Eq. (4), especially in the ISR energy range.

As we see, the shadow interpretation of the elastic scattering helps to understand many phenomena in the field and brings to the fore
the studying of the inelastic interaction intensity increasing with energy. Besides that, it gives some additional guidance for constructing
of adequate models of the elastic hadron-hadron scattering.

\section{Acknowledgements}
I am grateful to V.A. Petrov and I.M. Dremin for useful stimulating discussions and the critical remarks.

\end{document}